\documentclass[11pt]{article}
\usepackage{amsmath,amssymb,color,graphics,epsfig}

\usepackage{amsfonts}

\newcommand{\be}{\begin{equation}}
\newcommand{\ee}{\end{equation}}
\newcommand{\bea}{\setlength\arraycolsep{2pt} \begin{eqnarray}}
\newcommand{\eea}{\end{eqnarray}}
\newcommand{\nn}{\nonumber}

\def\ft#1#2{{\textstyle{\frac{\scriptstyle #1}{\scriptstyle #2} } }}
\def\fft#1#2{{\frac{#1}{#2}}}

\def\0{{\sst{(0)}}}
\def\1{{\sst{(1)}}}
\def\2{{\sst{(2)}}}
\def\3{{\sst{(3)}}}
\def\4{{\sst{(4)}}}
\def\5{{\sst{(5)}}}
\def\6{{\sst{(6)}}}
\def\7{{\sst{(7)}}}
\def\8{{\sst{(8)}}}
\def\sst#1{{\scriptscriptstyle #1}}

\begin{document}

\begin{flushright}
\end{flushright}

\vspace{25pt}
\begin{center}
{\large {\bf Note on Noether charge and holographic transports}}

\vspace{10pt}
 Zhong-Ying Fan

\vspace{10pt}
{ Center for Astrophysics, School of Physics and Electronic Engineering, \\
 Guangzhou University, Guangzhou 510006, China }\\
\smallskip

\vspace{40pt}

\underline{ABSTRACT}
\end{center}
We clarify the relation between the Noether charge associated to an arbitrary vector field and the equations of motions by revisiting Wald formalism. For a time-like Killing vector, aspects of the Noether charge suggest that it is dual to the heat current in the boundary for general holographic theories. For a space-like Killing vector, we interpret the Noether charge (at the transverse direction) as shear stress of the dual fluid so we can compute the ratio of shear viscosity to entropy density by simply using the infrared data on the black hole event horizon. We test the new method for Einstein gravity and Gauss-Bonnet gravity and find that it produces correct results for both cases even in the presence of additional matter fields.

\vfill {\footnotesize  Email: fanzhy@gzhu.edu.cn\,.}

\thispagestyle{empty}

\pagebreak

\tableofcontents
\addtocontents{toc}{\protect\setcounter{tocdepth}{2}}




\section{Introduction}

The AdS/CFT correspondence provides a new powerful method to compute the transport coefficients of strongly coupled systems that live on the boundary of asymptotically anti-de Sitter (AdS) space-times, including the ratio of shear viscosity to entropy density \cite{Buchel:2003tz,Buchel:2004qq,Benincasa:2006fu,Landsteiner:2007bd,Cremonini:2011iq,Iqbal:2008by,Cai:2008ph,Cai:2009zv,Brustein:2007jj,Kats:2007mq,shenker,
Natsuume:2010ky,Erdmenger:2010xm,Ovdat:2014ipa,Ge:2014aza,Liu:2015tqa,
Sadeghi:2015vaa,Parvizi:2017boc} and thermal-electric conductivities (for nice reviews, see \cite{Hartnoll:2009sz,Herzog:2009xv,Hartnoll:2016apf}). This is achieved by analysing small perturbations about the black holes background which describe the equilibrium state at finite temperature and chemical potential. In general, one considers perturbations with a time dependence of the form $e^{-i\omega t}$ and imposes ingoing boundary conditions on the black hole event horizon. Integrating out to infinity, one obtains the asymptotic solutions at AdS infinity. The retarded Green's functions of the boundary operators dual to the perturbations can be read off from the asymptotic solutions by using the holographic prescription \cite{sonsta}. The transport coefficients are related to the retarded Green's functions via the Kubo formula. For example, to compute the electric AC conductivity, one considers a time-dependent perturbation $\delta A_x(t\,,r)=a_x(r)e^{-i\omega t}$ and deduces the retarded Green's function $G^R_{J^x J^x}(\omega)$. The AC conductivity is given by $\sigma(\omega)=G^R_{J^xJ^x}(\omega)/\big(i\omega\big)$. By carefully taking the limit $\omega\rightarrow 0$, one can also obtain the DC conductivity (if the black hole background breaks translational invariance).

However, in some cases one may be more interested in direct calculations of the DC response. It was shown in \cite{shenker} that to compute shear viscosity to entropy ratio directly, one can analytically solve the linearized equation of motion for the transverse perturbation $\delta g_{xy}(t\,,r)$ at small $\omega$ approximation. This is successful for Einstein gravity with or without coupling to matter fields, even if one does not know the exact form of the black hole background \cite{Liu:2015tqa}. For pure Gauss-Bonnet gravity, it was found \cite{shenker} that the ratio receives corrections from the Gauss-Bonnet coupling and hence violates the conjectured universal bound (known as KSS bound) $\eta/S\geq 1/(4\pi)$ in \cite{Policastro:2001yc,KSS,KSS0}. A similar method was developed in \cite{Lucas:2015vna} for computing DC electric conductivity, which was solely expressed in terms of horizon data. Though the above results are remarkable, the method are somewhat tedious. It was first established in \cite{Donos:2014uba,Donos:2014cya} that if one instead considers perturbations with a linear time-dependence, one can calculate the electric, thermal and thermoelectric conductivities by simply using the near horizon solutions. This greatly simplifies the calculations of DC transport coefficients except for the shear viscosity. However, the method heavily depends on the construction of bulk radially conserved charges which correspond to the electric current and heat current in the boundary. In particular, for the heat current it is technically difficult according to its definition $Q_{\mathrm{heat}}^{x_i}=T^{tx_i}-\mu J_e^{x_i}$, where the boundary stress tensor $T^{ab}$ should be calculated by using holographic renormalization. For Einstein-Maxwell-Dilaton theories, the authors in \cite{Donos:2014cya} presented a nice result by constructing a radially conserved charge independent of the holographic stress tensor but for general cases the situation is not clear.

Quite recently, it was argued in \cite{Liu:2017kml} that the Noether charge associated to a time-like Killing vector field is dual to the holographic heat current. This was verified in \cite{Liu:2017kml} for Einstein gravity, Horndeski gravities and general Love-Lock gravities by calculating the heat current from holographic stress tensor independently. However, a shortcoming of the discussions in \cite{Liu:2017kml} (and also in \cite{Donos:2014cya}) is the connections of the radially conserved charges to the equations of motions are not clear. So one may doubt why the results there should be correct for general holographic theories, without calculating holographic stress tensor explicitly. In this paper, by carefully examining Wald formalism, we clarify the relation between the Noether charge associated to an arbitrary vector field and the Einstein equations of motions (see sec.\ref{sec2}). In particular, for either a time-like or a space-like Killing vector field, some components of the Noether charges will be radially conserved if the dual perturbations depend linearly on time. In fact, these charges turn out to be the integration constants of the linearized equations of motions which determine the dynamics of the perturbations. By analysing aspects of the Noether charge associated to a time-like Killing vector field, we argue that it provides an alternatively reasonable definition for the heat current in the boundary for general holographic theories.

Moreover, we consider the Noether charge associated to a space-like Killing vector field $\xi=\partial/\partial x$. We interpret the Noether charge $\mathcal{Q}^{ry}$ at the transverse direction as shear stress of the boundary fluid such that we can compute shear viscosity by simply using the infrared (IR) data on the black hole event horizon. This reproduces the celebrated result $\eta/S=1/(4\pi)$ for Einstein gravity even in the presence of matter fields. For Gauss-Bonnet gravity, we find that the ratio depends on the initial value of the scalar potential on the horizon in addition to the Gauss-Bonnet coupling. This is a new result that has never been found before. For pure Gauss-Bonnet gravity, it reproduces the correct result in \cite{shenker}. For general cases, in order to test the result, we rederive the shear viscosity to entropy ratio by using the conventional approach. By numerically solving the linearized equation of motion at small $\omega$ approximation, we find that our numerical result is perfectly matched with the above analytical result. This gives us strong confidence that our new method should be correct.

The paper is organized as follows. In section 2, we revisit the Wald formalism and clarify the relation between the Noether charge and the equations of motions. For a time-like Killing vector field, we study aspects of the Noether charge and argue that it is dual to the heat current of the boundary theory. In section 3, we study the Noether charge associated to a space-like Killing vector field. We interpret it as shear stress of the dual fluid and propose a new method to compute the shear viscosity to entropy ratio. We conclude in section 4.

\section{Noether charge and holographic heat current revisited}\label{sec2}
It was first developed by Wald \cite{wald1,wald2} that for a generic gravity theory with diffeomorphism invariance, the first law of thermodynamics for a stationary black hole can be systematically derived via the Noether charge associated to a time-like Killing vector field. The method is known as {\it Wald formalism} in literature. Recently, it has been applied to study the first laws of a variety of hairy black holes \cite{Papadimitriou:2005ii,Lu:2013ura,Liu:2013gja,Liu:2014tra,Liu:2014dva,Fan:2014ixa,Lu:2014maa,Fan:2014ala,Fan:2015yza,Chen:2016qks,Feng:2015oea,
Feng:2015wvb,Fan:2016jnz,Fan:2017bka}. In particular, it was established in \cite{Liu:2017kml} that the holographic heat current excited by transverse perturbations $\delta g_{tx_i}\,,\delta A_{x_i}$ with a linear time dependence was simply given by the Noether charge at asymptotic infinity (see Eq.(\ref{heat})). To begin our story, let us first review how Noether charge is introduced in Wald formalism.

Variation of the action with respect to dynamical fields (the metric and the matter fields), one finds
\be \delta \Big(\sqrt{-g}\mathcal{L} \Big)=\sqrt{-g}\Big(E_\Phi \delta \Phi+\nabla_\mu J^{\mu}\Big) \,,\ee
where $\Phi$ collectively denotes all the dynamical fields (without confusion, tensor indices have been omitted for sake of simplicity) and $E_\Phi=0$ are equations of motions.
Note that the current $J^\mu$ depends linearly on the variation of the dynamical fields $\delta \Phi$ and it is uniquely fixed up to a total derive term in the action. Given the current $J^\mu$, one can define a current 1-form and a current (n-1)-form as follows
\be J_{(1)}:=J_\mu dx^\mu,\qquad \Theta_{(n-1)}:=*J_{(1)} \label{joneform}\,.\ee
One can further define a Noether current $(n-1)$-form as:
\be J_{(n-1)}:=\Theta_{(n-1)}-i_\xi\cdot *\mathcal{L} \label{noetherj}\ee
where $i_\xi\cdot$ denotes the contraction of $\xi$ to the first index of the tensor it acted upon. Wald first shows that once the equations of motions are satisfied, one has
\be dJ_{(n-1)}=-E_{(n)}\delta \Phi= 0 \label{dnoetherj}\,,\ee
where $E_{(n)}$ denotes the $n$-form equations of motions. Thus, one can define a Noether charge (n-2)-form as
\be J_{(n-1)}:= dQ_{(n-2)}+E.O.M \label{noethercharge}\,,\ee
where $E.O.M$ denotes the terms proportional to the equations of motions. Then using the Noether charge 2-form $Q_{(n-2)}=*\mathcal{Q}_{(2)}$ and taking the dual form of the equation (\ref{noetherj}), one finds
\be *d*\mathcal{Q}_{(2)}=*J_{(n-1)}=*\Theta_{(n-1)}-*\Big(\xi_{(1)}\cdot *\mathcal{L} \Big) \,,\label{central1}\ee
where we have dropped the terms of equations of motions. This gives rise to the main result of \cite{Liu:2017kml}
\be \nabla_\nu \mathcal{Q}^{\mu\nu}=J^\mu-\xi^\mu\mathcal{L} \,.\label{central2}\ee
It should be emphasized that the above Noether charge is defined for any vector field $\xi$ that is not limited to a time-like Killing vector field. This is important for us to study its application to holographic transports. When $\xi$ is a Killing vector field of the space-time, one has $\delta \Phi=L_\xi \Phi=0$ such that $J^\mu(\delta\Phi)=0$ because the current $J^\mu$ depends linearly on $\delta \Phi$.
 In particular, for a static AdS planar black hole, the Noether charge $\mathcal{Q}^{rx_i}$ associated to a time-like Killing vector $\xi=\partial/\partial t$ is radially conserved, namely
\be \partial_r\Big(\sqrt{-g}\,\mathcal{Q}^{r x^i} \Big)=0 \,.\ee
With this observation, it was proposed in \cite{Liu:2017kml} that the above radially conserved Noether charge is nothing else but simply the holographic heat current $Q_{\mathrm{heat}}^{x_i}=T^{tx_i}-\mu J^{x_i}_e$, where $T^{ab}$ is the boundary stress tensor, $\mu$ is the chemical potential and $J_e^{x_i}$ is the electric current. One has
\be Q_{\mathrm{heat}}^{x_i}=-\sqrt{-g}\,\mathcal{Q}^{rx_i}\Big|_{\mathrm{boundary}}  \,.\label{heat}\ee
This conjecture has been tested for Einstein gravity, Horndeski gravities and general Lovelock gravities in \cite{Liu:2017kml} by computing the heat current from both the Noether charge and holographic stress tensor independently. In spite of that the results in \cite{Liu:2017kml} look very nice, one may doubt why the formula (\ref{heat}) should be correct for general holographic theories.

To clarify this, we shall first recover the E.O.M terms in the equation (\ref{central2}). To be concrete, we will focus on a generic gravity theory
$\mathcal{L}_{grav}(g_{\mu\nu}\,;R_{\mu\nu\rho\sigma})$ which is minimally coupled to a Maxwell field and a scalar field as
\be\label{lagrangian1} \mathcal{L}_{tot}=\kappa\,\mathcal{L}_{grav}(g_{\mu\nu}\,;R_{\mu\nu\rho\sigma})-\ft 14 Z(\phi)F^2-\ft 12 \big(\partial \phi\big)^2-V(\phi) \,,\ee
where the gravity coupling constant $\kappa=1/(16\pi G_N)$ will be set to unity throughout this paper. For pure gravity, we find
\bea\label{gravity} &&\nabla_\nu \mathcal{Q}^{\mu\nu}_{grav}=J^{\mu}_{grav}-\xi^\mu \mathcal{L}_{grav}-2\xi_\nu\, E^{\mu\nu}\,,\nn\\
&&\mathcal{Q}^{\mu\nu}_{grav}=-\Big( 2E^{\mu\nu\rho\sigma}\nabla_\rho \xi_\sigma+4\xi_\rho \nabla_\sigma E^{\mu\nu\rho\sigma} \Big)\,,\nn\\
&&J^\mu_{grav} =2\Big(E^{\sigma\alpha\beta\mu}\nabla_\sigma \delta g_{\alpha\beta}- \nabla_\sigma E^{\sigma\alpha\beta\mu}\delta g_{\alpha\beta}  \Big)\,,\nn\\
&&E_{\mu\nu}=E_{\mu\sigma\alpha\beta}R_{\nu}^{\,\,\,\sigma\alpha\beta}+2\nabla^\lambda \nabla^\rho E_{\mu\lambda\nu\rho}-\ft 12 g_{\mu\nu}\mathcal{L}_{grav}\,,
\eea
where $E^{\mu\nu\rho\sigma}\equiv \partial \mathcal{L}/\partial R_{\mu\nu\rho\sigma}$. For the Maxwell field, we find
\bea\label{maxwell}
&&\nabla_\nu \mathcal{Q}^{\mu\nu}_A=J^\mu_A-\xi^\mu \mathcal{L}_A+2\xi_\nu\, T^{\mu\nu}_A-E^\nu A_\sigma \xi^\sigma\,,\nn\\
&&\mathcal{Q}^{\mu\nu}_A=-F^{\mu\nu}A_\sigma \xi^\sigma\,,\quad J^\mu_{A}=-F^{\mu\nu}\delta A_\nu\,,\quad E^\mu=\nabla_\nu F^{\mu\nu}\,,
\eea
whilst for the scalar field $\mathcal{Q}^{\mu\nu}_\phi=0$ and
\be\label{scalar} 0=J^\mu_\phi-\xi^\mu \mathcal{L}_\phi+2\xi_\nu\, T^{\mu\nu}_\phi\,,\quad J^\mu_{\phi}=-\nabla^\mu\phi\, \delta \phi  \,.\ee
Combining the above results, we deduce
\be\label{main} \nabla_\nu \mathcal{Q}^{\mu\nu}_{tot}=J^\mu_{tot}-\xi^\mu \mathcal{L}_{tot}-2\xi_\nu\, H^{\mu\nu}-E^\mu A_\sigma \xi^\sigma\,.\ee
Here $H^{\mu\nu}\equiv E^{\mu\nu}-T^{\mu\nu}_A-T^{\mu\nu}_\phi$ are Einstein equations. It should be emphasized that the above result is valid as well when there are non-minimal couplings between the metric and the matter fields, with the Noether charge and the equations of motions including extra terms associated
to the non-minimal couplings. We refer the readers to literatures \cite{Chen:2016qks,Feng:2015oea,Feng:2015wvb,Fan:2016jnz,Fan:2017bka} for more details on this point. It becomes clear that the Noether charge $\mathcal{Q}^{rx_i}$ is closely related to the dynamical equations of the transverse perturbations $(\delta g_{tx_i}\,,\delta A_{x_i})$. We find that at linear order\footnote{For general time dependent perturbations, $J^\mu$ is does not vanish because of $J^\mu\sim \partial_t^2 \delta g_{tx_i}$.}
\be 0=\nabla_r \mathcal{Q}^{r x_i}_{tot}=2g_{tt}\, H^{tx_i}+A_t E^{x_i}  \,.\label{heateom}\ee
Hence, when integrating the r.h.s of the equation by part, one finds the charge $\sqrt{-g}\mathcal{Q}^{rx_i}$ is an integration constant of the linearized equations of motions. This is very similar to the case of electric current $J^{x_i}_e$ which satisfies
\be J^{x_i}_e=-\sqrt{-g}\, F^{rx_i}\Big|_{\mathrm{boundary}}\,,\quad 0=\nabla_r F^{r x_i}=E^{x_i} \,.\ee
As a matter of fact, the Noether charge may be viewed as an alternative but more convenient definition for holographic heat current of the boundary theory. The consistency between this definition and the original one which highly depends on the boundary stress energy tensor has been checked in a case-by-case basis in \cite{Donos:2014cya,Liu:2017kml} for certain classes of AdS gravities.

However, without a solid proof, we shall try to give some further supports for the above argument for general holographic theories. We consider a generally electrically charged static solution
\be\label{static} ds^2=-h(r) dt^2+dr^2/f(r)+r^2 dx^i dx^i\,,\quad A=A_t(r) dt\,,\quad \phi=\phi(r) \,.\ee
At the time direction of the equation (\ref{main}), we find a modified radially conserved charge
\bea c&=&-\sqrt{-g}\,\mathcal{Q}^{rt}+\int_{r_0}^r \mathrm{d} r\sqrt{-g}\,\mathcal{L}\,,\nn\\
  &=&\sqrt{h f}\Big(\fft{h'}{h}-\fft{2}{r} \Big)r^{n-2}-Q_e A_t\,.
\label{chargec}\eea
Evaluating $c$ on the horizon yields
\be c=T\,S \,.\ee
This is reminiscent of the known result \cite{wald1,wald2}
\be \fft{1}{16\pi G_N}\int Q_{(n-2)}=T\,S \,.\ee
Note that for convenience we have set the volume factor of the codimension-2 space $\omega_{n-2}\equiv \int_{\Sigma_{n-2}} dx_1dx_2\cdots dx_{n-2}$ to unity so that all extensive quantities in this paper should be understood as those of densities. On the other hand, at asymptotic infinity the metric functions and the matter fields behave as\footnote{Depending on the mass square of the scalar field, there could be some slower fall-off modes than the condensate of massless gravitons in the metric function $f$ but this can not happen in the metric function $h$. The absence of these intervening terms in the asymptotic behavior of $h$ can be deduced using the first order equation (\ref{chargec}). }
\bea\label{asympt}
&&h=g^2r^2-\fft{\mu_g}{r^{n-3}}+\cdots\,,\quad f=g^2r^2+\cdots\,,\nn\\
&&A_t=\mu-\fft{Q_e}{r^{n-3}}+\cdots\,,\quad \phi=\fft{\phi_-}{r^{\Delta_-}}+\fft{\phi_+}{r^{\Delta_+}}+\cdots \,,\eea
where $g=1/\ell$ is the inverse of AdS radius and $\Delta_\pm=\ft 12(n-1\pm \nu)\,,\nu=\sqrt{4 m^2 \ell^2+(n-1)^2}$, $m^2$ is the mass square of the scalar field. $\mu_g$ is an integration constant associated to the condensate of massless gravitons. The black hole mass is given by
\be M=\fft{(n-2)\mu_g}{16\pi G_N} \,,\ee
and the electric charge $Q_e$ carried by the black hole is defined as
\be Q_e=\fft{1}{16\pi G_N}\int *F \,. \ee
Then evaluating $c$ at infinity yields
\be c=(n-1)\mu_g+\mu\, Q_e =\ft{n-1}{n-2}M+\mu\,Q_e\,.\ee
So we deduce
\be M=\fft{n-2}{n-1}\Big(T\,S+\mu\,Q_e \Big) \,.\label{smarr}\ee
This is exactly the generalized Smarr relation \cite{Liu:2015tqa} which is associated to the scaling symmetry of AdS planar black holes
\be r\rightarrow \lambda\, r\,,\quad \big(t\,,x_i\big)\rightarrow \lambda^{-1}\big(t\,,x_i\big)\,,\quad \big(h\,,f \big)\rightarrow \lambda^2 \big(h\,,f\big)\,,\quad A_t\rightarrow \lambda\, A_t \,.\ee
The radially conserved charge $c$ is nothing else but the Noether charge of the above scaling symmetry. It was further established in \cite{Liu:2015tqa} that the generalized Smarr relation plays an indispensable role in the derivation of shear viscosity to entropy ratio in the conventional approach. We will come to this point in the next section. Here it is worth emphasizing that
\be -\sqrt{-g}\,\mathcal{Q}^{rt}=T\,S \,,\ee
may be interpreted as thermal energy of the boundary theory according to the Clausius relation. Recall the relation (\ref{heat}), the covariant thermal current may be identified as
\be\label{covariantheat} J_Q^a=-\sqrt{-g}\,\mathcal{Q}^{r a}\Big|_{\mathrm{boundary}} \,.\ee
Then the conservation of the thermal energy in the boundary $\partial_a J^a_Q=0$ is a natural result of the bulk identity $*d*\mathcal{Q}_{(2)}=0$ at the radial direction. This is very similar to the covariant electric current in the boundary,
\be J^a_e=-\sqrt{-g}\, F^{r a}\Big|_{\mathrm{boundary}} \,,\ee
where $J_e^t$ is the charge density and $\partial_a J^a_e=0$ owing to $*d* F=0$. Hence, in addition to the relation (\ref{heateom}), the above comparison with the electric current further supports our argument that the Noether charge of a time-like Killing vector field may be viewed as an alternatively reasonable definition for holographic heat current.

\section{Shear viscosity to entropy ratio}
Observing that our main result (\ref{main}) connects the Noether charge of local diffeomorphism invariance and the Einstein's equations of motions, we are able to propose a new method to compute shear viscosity to entropy ratio by taking $\xi=\partial/\partial x_1$. Before moving to details, let us first review the conventional approach adopted in the literature \cite{shenker,Liu:2015tqa}. One considers a general time dependent perturbation $\delta g_{x_1x_2}(t\,,r)=g_{x_1x_1}\Psi(t\,,r)=g_{x_1x_1} \psi(r)e^{-i\omega t}$ and solves the linearized Einstein equation $H_{x_1x_2}$ to relevant order in small frequency limit, subject to in-going boundary conditions on the horizon. Evaluating the action at quadratic order, one can deduce the shear viscosity by making use of the asymptotic solutions at infinity. For Einstein gravity, the result is of the form \cite{Liu:2015tqa}
\be \eta\sim \big(\ft{n-1}{n-2}M-\mu\,Q_e \big)/T \,,\ee
up to a constant coefficient that depends on the parameters of the theories. One further adopts the generalized Smarr relation (\ref{smarr}) and finds that $\eta/S$ is an universal constant which does not depend on any details of the solutions\footnote{For Einstein gravity, $\eta/S=1/4\pi$ whilst for higher curvature gravity such as Gauss-Bonnet gravity the ratio receives corrections from higher order gravitational coupling constants.}. This is a remarkable result which motives us to search a new method to compute shear viscosity by simply using the near horizon solutions instead of the asymptotic solutions.

\begin{figure}[ht]
\centering
\includegraphics[width=230pt]{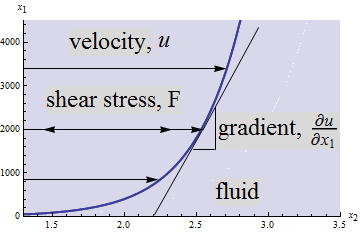}
\caption{{\it The cartoon picture of a 2-dimensional fluid. The relative motion of adjacent layers generates the shear stress, which is proportional to the gradient of the fluid velocity. }}
\label{fig0}\end{figure}

Moving back to our result (\ref{main}), we find that for a space-like Killing vector $\xi=\partial/\partial x_1$, the current $J^\mu$ vanishes as well when the transverse perturbation $\delta g_{x_1x_2}$ depends linearly on time. Thus, at the transverse direction such as $x_2$, we find
\be 0=\nabla_r \mathcal{Q}^{rx_2}=2g_{x_1x_1}H^{x_1x_2}  \,.\ee
In other words, the charge $\sqrt{-g}Q^{rx_2}$ associated to $\xi=\partial/\partial x_1$ turns out to be an integration constant of the Einstein equation $H^{x_1x_2}$ that determines the dynamics of the transverse perturbation. Since $\sqrt{-g}\mathcal{Q}^{rx_2}$ is radially conserved, we can easily calculate it from simply the near horizon solutions. This is in the same spirit of the calculations of thermal-electric conductivities in \cite{Donos:2014uba,Donos:2014cya}. One remaining question is what the physical meaning of $\mathcal{Q}^{rx_2}$ in the boundary theory and whether this can reproduce correct results for shear viscosity to entropy ratio in comparable with the conventional approach. To explain this, we consider a 2-dimensional fluid system in the boundary (see Fig.\ref{fig0}). The flow is moving along $x_2$ axis and has a velocity $u=u(x_1)$, which depends only on the perpendicular axis. Along the $x_1$ axis, the fluid can be thought of as infinite layers which move parallel to each other with different speeds.
This effectively generates a shear stress $F^{x_2}$ between the adjacent layers. The viscosity of the fluid is defined by its resistance to the shearing flows, namely
\be F^{x_2}\equiv \eta\, \ft{\partial u}{\partial x_1}  \,.\label{sheardefinition}\ee
This is an Ohm-like relation that is similar to the definition of thermal-electric conductivities. Thus, we propose
\be F^{x_2}=-\sqrt{-g}\,\mathcal{Q}^{rx_2}\Big|_{\mathrm{boundary}} \,. \label{shearstress}\ee
We will show that this indeed reproduces correct results for shear viscosity to entropy ratio by considering a transverse perturbation $\delta g_{x_1\,,x_2}(t\,,r)$ with a linear time dependence.

\subsection{Einstein gravity}
To be specific, let us first consider Einstein gravity by setting $\mathcal{L}_{grav}=R$ in the Lagrangian density (\ref{lagrangian1}). We have
\bea\label{gravity2} &&\nabla_\nu \mathcal{Q}^{\mu\nu}_{grav}=J^{\mu}_{grav}-\xi^\mu R-2\xi_\nu\, G^{\mu\nu}\,,\nn\\
&&\mathcal{Q}^{\mu\nu}_{grav}=-2\nabla^{[\mu} \xi^{\nu]}\,,\quad J^\mu_{grav} =g^{\mu\nu}g^{\rho\sigma}\big(\triangledown_\sigma \delta g_{\nu\rho}-\triangledown_\nu \delta g_{\rho\sigma} \big)\,,
\eea
The scalar potential $V(\phi)$ has a small $\phi$ expansion as
\be V=-(n-1)(n-2)\ell^{-2}+\ft 12 m^2 \phi^2+\gamma_3\phi^3+\gamma_4\phi^4+\cdots \,,\ee
where $\ell$ denotes bare AdS radius. The static background still takes the form of (\ref{static}). At $x_1$ direction, the equation (\ref{main}) gives rise to a modified radially conserved charge
\be c'=-\sqrt{-g}\,\mathcal{Q}^{rx_1}+\int_{r_0}^r \mathrm{d}r \sqrt{-g}\,\mathcal{L} \,. \ee
However, by substituting into the background solutions we find that $c'$ is trivial: $c'=0$. Next, we perturb the background as
\be\label{transverse} \delta g_{x_1x_2}(t\,,r)=g_{x_1x_1}\zeta\, t+h_{x_1x_2}(r) \,,\ee
where $\zeta$ is identified as the gradient of the fluid velocity along the $x_1$ direction, namely $\zeta\equiv \partial u/\partial x_1$. According to our conjecture (\ref{shearstress}), the shear stress is given by
\be F^{x_2}=-\sqrt{-g}\,\mathcal{Q}^{rx_2}=r^{n-2} \sqrt{hf}\, \partial_r\big(r^{-2}h_{x_1x_2}\big) \,.\ee
To express the shear stress in terms of horizon data, we develop Taylor expansions for the metric functions and matter fields in the near horizon region
\bea\label{horizon}
&&h=h_1(r-r_0)+h_2(r-r_0)^2+\cdots\,,\quad f=f_1(r-r_0)+f_2(r-r_0)^2+\cdots \,,\nn\\
&&\phi=\phi_0+\phi_1(r-r_0)+\phi_2(r-r_0)^2+\cdots\,,\quad A_t=a_1(r-r_0)+a_2(r-r_0)^2+\cdots\,,
\eea
and impose in-going boundary condition for the perturbation
\be\label{ingoing} h_{x_1x_2}=\fft{\zeta\,r^2_0 }{4\pi T} \log{\big( r-r_0 \big)}+\cdots \,.\ee
Then we deduce
\be F=\zeta r_0^{n-2}=\zeta \fft{S}{4\pi} \,,\ee
where we have used
\be T=\fft{\sqrt{h_1f_1}}{4\pi}\,,\quad S=4\pi\, r_0^{n-2} \,.\ee
By definition, the shear viscosity is given by
\be \eta=\fft{F}{\zeta}=\fft{S}{4\pi} \,.\ee
This reproduces the celebrated result
\be \fft{\eta}{S}=\fft{1}{4\pi} \,.\ee
It is easily seen that our new method is more easier than the conventional one. In fact, it is particularly powerful in the calculations of shear viscosity for higher curvature/derivative gravities, in which cases analytical results in general cannot be derived by using the conventional approach, as will be shown later.

\subsection{Gauss-Bonnet gravity}
\subsubsection{New method}
We continue to study shear viscosity to entropy ratio in Gauss-Bonnet gravity. The gravitational Lagrangian density is given by
\be \mathcal{L}=R+\alpha \big(R^2-4R^2_{\mu\nu}+R^2_{\mu\nu\rho\sigma} \big) \,.\ee
The current $J^\mu$ and the Noether charge are given by \cite{Fan:2014ala,Chen:2016qks}
\bea
\mathcal{Q}^{\mu\nu} &=& -2\Big(\nabla^{[\mu}\xi^{\nu]}
+2\alpha\,\big(R\,\nabla^{[\mu}\xi^{\nu]}-4 R^{\sigma[\mu}\nabla_\sigma \xi^{\nu]}+ R^{\mu\nu\sigma\rho}\nabla_\sigma \xi_\rho \big)\Big)\,,\nn\\
J^\mu&=&
\Big(G^{\mu\nu\rho\lambda}+2\alpha\, \big(R\, G^{\mu\nu\rho\lambda}-2 T^{\mu\nu\rho\lambda}+2 R^{\mu\rho\lambda\nu} \big)\Big)\, \nabla_\nu \delta g_{\rho\lambda}\nn\\
&&-2\alpha\, \big(G^{\mu\nu\rho\lambda}\nabla_\nu R-2 \nabla_\nu T^{\nu\mu\rho\lambda}+2 \nabla_\nu R^{\mu\rho\lambda\nu}\big)\, \delta g_{\rho\lambda}\,,
\eea
where
\bea G^{\mu\nu\rho\sigma} &=&\ft 12 (g^{\mu\rho}g^{\nu\sigma}+g^{\mu\sigma}g^{\nu\rho}) - g^{\mu\nu}g^{\rho\sigma},\cr
T^{\mu\nu\rho\lambda} &=& g^{\mu\rho}R^{\nu\lambda} +g^{\mu\lambda}R^{\nu\rho}-g^{\mu\nu}R^{\rho\lambda}-
g^{\rho\lambda}R^{\mu\nu}\,.\label{gt}
\eea
Evaluating the Noether charge for the static solution (\ref{static}) with the transverse perturbation (\ref{transverse}) yields
\be F=-\sqrt{-g}\,\mathcal{Q}^{rx_2}=\Big(1-2\alpha\,(n-4)\big(\ft{h'}{h}+\ft{n-5}{r} \big)\ft{f}{r} \Big)\,r^{n-2}\sqrt{hf}\,\partial_r\big(r^{-2}h_{x_1x_2} \big) \,.\ee
By plugging the near horizon solutions (\ref{horizon}-\ref{ingoing}) into this equation, we obtain
\be F=\zeta\Big( 1-2\alpha\,(n-4)f_1/r_0 \Big)r_0^{n-2} \,.\ee
It follows that the shear viscosity to entropy ratio is given by
\be \fft{\eta}{S}=\fft{1}{4\pi}\Big( 1-2\alpha\,(n-4)f_1/r_0 \Big) \,.\ee
The result depends on the Gauss-Bonnet coupling as well as the horizon data $f'(r_0)/r_0$. By expanding the background equations of motions in the near horizon region, we find that the near horizon solutions are characterized by four independent parameters $(h_1\,,r_0\,,\phi_0\,,a_1)$, where $h_1$ is a trivial parameter associated with the scaling symmetry of the time coordinate. All the rest of the coefficients can be solved in terms of functions of these four parameters. In particular, we find
\be f_1=-\fft{V(\phi_0)}{n-2}\,r_0 \,.\ee
So we deduce
\be \fft{\eta}{S}=\fft{1}{4\pi}\Big( 1+\ft{2(n-4)}{n-2}\,\alpha\,V(\phi_0)\Big) \,.\label{shearGB}\ee
It is interesting to note that the above result receives corrections from dimensionless parameters of the theories $(\alpha \ell^{-2}\,,\alpha m^2\,,\alpha \gamma_3\,,\alpha\gamma_4\,,\cdots)$ as well as the initial data of the scalar field on the horizon. This is a new result that has never been found before.
For pure Gauss-Bonnet gravity, it reproduces the known result in \cite{shenker}
\be \fft{\eta}{S}=\fft{1}{4\pi}\Big( 1-2(n-1)(n-4)\,\alpha\,\ell^{-2}\Big) \,.\label{shearGBpure}\ee
However, in the presence of matter fields, it is of great difficult to derive an analytical expression such as (\ref{shearGB}) by using the conventional method for Gauss-Bonnet gravity.

By analysing causality violation for Gauss-Bonnet black hole\footnote{The analysis in \cite{Brigante:2008gz} was performed for $n=5$ dimensional Gauss-Bonnet black hole. It was found that when $\alpha\ell^{-2}>9/200$, the theory violates causality and hence is inconsistent. This gives rise to a lower bound on viscosity: $\eta/S\geq 16/100\pi$.}, it was argued in \cite{Brigante:2008gz} that the r.h.s of (\ref{shearGBpure}) violates KSS bound for $\alpha>0$, giving rise to a modified lower bound on viscosity for a class of conformal field theories with Gauss-Bonnet gravity dual. However, our new result (\ref{shearGB}) implies that the lower bound proposed in \cite{Brigante:2008gz} may be violated as well in the presence of a scalar field. Indeed, in the Appendix, by following the analysis in \cite{Brigante:2008gz}, we explore the viscosity bound and causality violation for Gauss-Bonnet black holes with scalar hair. Our numerical results suggest that with a scalar field, the viscosity still has a lower bound which is however decreased further by the scalar hair.


\subsubsection{Conventional method}

To give a cross check for our result (\ref{shearGB}) in the presence of matter fields, we shall calculate the shear viscosity to entropy ratio by making use of the conventional approach. We follow closely the procedure established in \cite{shenker}. For later convenience, we focus on the $n=5$ dimension and turn off the Maxwell field. Considering a transverse perturbation $\delta g_{x_1x_2}(t\,,r)=r^2\psi(r)e^{-i\omega t}$, we derive the perturbative Einstein equation
\bea\label{exy}
0&=&hf\Big(1-\ft{2\alpha\,f h'}{r h} \Big)\,\psi''+\Big(\ft 12\big(h f\big)'+\ft{3hf}{r}\nn\\
&&-\ft{\alpha f^2 h'}{r}\big(\ft{2h''}{h'}-\ft{h'}{h}+\ft{3f'}{f}+\ft{4}{r} \big) \Big)\,\psi'+\big(1-\ft{2\alpha f'}{r} \big)\,\omega^2\psi\,,
\eea
where we have used the background equations of motions to simplify the result. On the event horizon, we impose ingoing boundary condition
\be \psi(r)\propto \mathrm{exp}{\Big[-\fft{i \omega}{4\pi T}\log{(r-r_0)}\Big]} \,.\ee
At asymptotic infinity, the metric functions and the scalar field behave as
\be\label{asymp2}
h=\widetilde{g}^2r^2-\fft{\mu_g}{r^2}+\cdots\,,\quad f=\widetilde{g}^2r^2+\cdots\,,\quad \phi=\fft{\phi_-}{r^{\Delta_-}}+\fft{\phi_+}{r^{\Delta_+}}+\cdots\,,
\ee
where $\widetilde{g}$ denotes the inverse of effective AdS radius. It is related to the bare AdS radius $g=1/\ell$ as
\be g^{2}=\widetilde{g}^2\Big(1-(n-3)(n-4)\alpha\widetilde{g}^2 \Big) \,.\ee
The function $\psi(r)$ behaves asymptotically as
\be \psi(r)=J+\fft{\mathcal{O}}{r^4}+\cdots \,.\ee
Here $J$ and $\mathcal{O}$ are dual to the source and vacuum expectation value of the boundary operator respectively.
 In order to extract the shear viscosity using the Kubo formula, we need only know $\psi$ up to linear order in $\omega$ so we may seek the solution of the equation (\ref{exy}) in the small $\omega$ approximation. As a matter of fact, it is convenient to make an ansatz of
\be \psi(r)= J\,\mathrm{exp}{\Big[-\fft{i \omega}{4\pi T}\log{\Big(\ft{h(r)}{\widetilde{g}^2r^2}\Big)}\Big]}\Big(1-i\omega\, T\, U(r) \Big)\,.\ee
Then substituting the above ansatz into the equation (\ref{exy}) and expanding it to linear order in $\omega$, we arrive at a fairly complicated ordinary differential equation for the function $U$
\be\label{ueq} U''+P_1(r)U'+P_2(r)=0 \,,\ee
where the functions $P_1\,,P_2$ have lengthy expressions which we do not find instructive to present. It is quite difficult to solve this equation analytically except for the case of Einstein gravity corresponding to $\alpha=0$. So later on, we will adopt numerical method to solve the equation and deduce the shear viscosity to entropy ratio. We demand that $U$ is a regular function on the horizon whilst at infinity, it behaves as
\be U(r)=\fft{u_+}{4\pi\widetilde{g}^2 r^4}+\cdots \,.\ee
Thus, the asymptotic behavior of the function $\psi$ is given by
\be\label{asymp3} \psi=J\,\Big[1+\fft{i\omega\big(\mu_g-u_+ \big)}{4\pi \widetilde{g}^2\,T r^4}+\cdots \Big] \,.\ee

In order to deduce the shear viscosity using the prescription in \cite{sonsta}, we shall compute the action to quadratic order, including Gibbons-Hawking boundary term. This is easily done by making use of the linearized equation of motion. We find
\be S_2=\int \mathrm{d}^5x \Big[\partial_r\big(K_1\Psi\Psi'+\ft 12 K_2\Psi^2 \big)+\partial_t\big(K_3 \Psi\dot \Psi \big)-\Psi\, H_{x_1x_2} \Big] \,,\ee
where $K_1\,,K_2\,,K_3$ are three functions of $r$. For our purpose, the only relevant term is $K_1$, given by
\be K_1=-\ft 12 r^3\sqrt{hf}\Big(1-\ft{2\alpha f h'}{r h} \Big) \,.\ee
By plugging the asymptotic solutions (\ref{asymp2}) and (\ref{asymp3}) into the quadratic action, we find
\be S_2=\fft{i\omega J^2}{2\pi T}\big(\mu_g-u_+ \big)\big(1-4\alpha \widetilde{g}^2 \big) \,.\ee
It follows that the shear viscosity to entropy ratio is given by
\be\label{shearnum} \fft{\eta}{S}=\fft{1}{4\pi}\Big(1-\fft{4\pi \widetilde{g}^2u_+}{\mu_g}\Big) \,.\ee
Here we have adopted the generalized Smarr relation of Gauss-Bonnet black holes \cite{Liu:2015tqa}
\be (n-1)\Big[1-2\alpha(n-3)(n-4)\widetilde{g}^2 \Big]\mu_g=T S \,.\ee
\begin{figure}[ht]
\includegraphics[width=210pt]{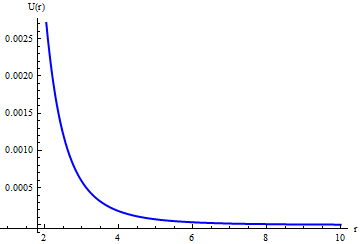}
\includegraphics[width=210pt]{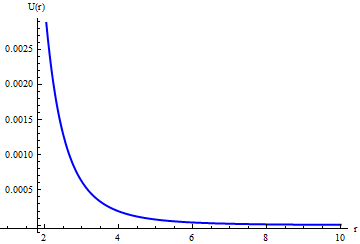}
\caption{{\it The plots of $U(r)$ for $m^2=1/4$ (left panel) and $m^2=-4$ (right panel). We have set $\ell=1\,,\alpha=1/16\,,T=1/10\,,\phi_0=1/2$.}}
\label{fig1}\end{figure}
It is worth emphasizing that the result (\ref{shearnum}) depends on details of the asymptotic solutions. Since we are not able to solve the function $U(r)$ analytically, we do not know whether the result (\ref{shearnum}) can be expressed solely in terms of the IR data on the black hole event horizon such as (\ref{shearGB}). Hence, by solving the equation (\ref{ueq}) numerically and comparing the result (\ref{shearnum}) with the analytical result (\ref{shearGB}), we can give a highly non-trivial cross check for our new approach.

To perform numerical calculations, we focus on a free massive scalar field. The Breitenlohner-Freedman (BF) bound is $m^2_{BF}=-\ft 14(n-1)^2\widetilde{g}^2$. Using the scaling symmetry
\be r\rightarrow \lambda\,r\,,\quad (t\,,x_i)\rightarrow \lambda^{-1}(t\,,x_i)\,,\quad (h\,,f)\rightarrow \lambda\,(h\,,f) \,,\ee
we can set the horizon radius to unity. Then there are two remaining parameters on the horizon which we may take to be the temperature $T$ and $\phi_0$ (Note that we have turned off the Maxwell field, so the background solution has one less independent parameters on the event horizon). By properly specifying these initial data on the horizon as well as the coupling constants of the theories, we can numerically solve the function $U(r)$, together with the metric functions and the scalar field.
\begin{figure}[ht]
\includegraphics[width=210pt]{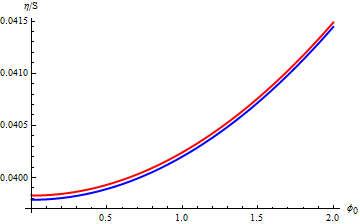}
\includegraphics[width=210pt]{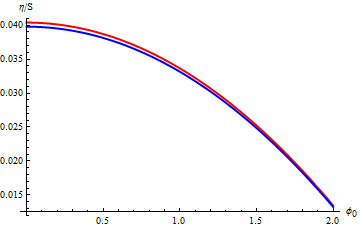}
\caption{{\it The shear viscosity to entropy ratio as a function of $\phi_0$ for $m^2=1/4$ (left panel) and $m^2=-4$ (right panel), respectively. The red curve is (\ref{shearGB}) and the blue curve is (\ref{shearnum}). We have slightly scaled (\ref{shearGB}) to have a nice presentation. We have set $\ell=1\,,\alpha=1/16$.}}
\label{fig2}\end{figure}
In Fig.\ref{fig1}, we plot the function $U(r)$ for both mass square $m^2=1/4$ and $m^2=-4$, respectively. The falloff mode $u_+$ can be easily read off from the numerical solutions at asymptotic infinity. We find that at sufficiently high precision, the shear viscosity to entropy ratio does not explicitly depend on on the temperature for any given $\phi_0$. This is consistent with our result (\ref{shearGB}). In Fig.\ref{fig2}, we plot $\eta/S$ for both (\ref{shearGB}) and (\ref{shearnum}) as a function of $\phi_0$, corresponding to the red curve and blue curve respectively. To have a nice presentation, we have slightly scaled (\ref{shearGB}). We find that the numerical result for (\ref{shearnum}) matches well with the analytical result (\ref{shearGB}).

In Table.\ref{sheartable1} and Table.\ref{sheartable2}, we present the arguments of $\eta/S$ for both (\ref{shearGB}) and (\ref{shearnum}) for several values of $\phi_0$. The two results just have tiny differences which are of order $10^{-5}\sim 10^{-6}$. Thus, we can safely conclude that our numerical results from the conventional method verify the analytical result (\ref{shearGB}). This gives us strong confidence that our new approach to compute the shear viscosity is correct. In fact, as already shown it is more convenient and more powerful than the conventional method for higher curvature/derivative gravities.
\begin{table}[h]
\centering
\begin{tabular}{|c|c|c|c|c|c|}
  \hline
  $\phi_0$ & 0.1  & 0.5  & 0.9  & 1.3 & 1.7   \\ \hline
  Eq.(\ref{shearGB}) & 0.03979288  & 0.03989235   & 0.04012445  & 0.04048918 & 0.04098654 \\ \hline
  Eq.(\ref{shearnum}) & 0.03979261  & 0.03989201  & 0.04012404  & 0.04048844 & 0.04098578  \\ \hline
\end{tabular}
\caption{The shear viscosity to entropy ratio for $m^2=1/4$. We have set $\ell=1\,,\alpha=1/16$.}
\label{sheartable1}
\end{table}
\begin{table}[h]
\centering
\begin{tabular}{|c|c|c|c|c|c|}
  \hline
  $\phi_0$ & 0.1  & 0.5  & 0.9  & 1.3 & 1.7   \\ \hline
  Eq.(\ref{shearGB}) & 0.03972242  & 0.03813087   & 0.03441726  & 0.02858158 & 0.02062383 \\ \hline
  Eq.(\ref{shearnum}) & 0.03972161  & 0.03812956  & 0.03441507  & 0.02857791 & 0.02061803  \\ \hline
\end{tabular}
\caption{The shear viscosity to entropy ratio for $m^2=-4$. We have set $\ell=1\,,\alpha=1/16$.}
\label{sheartable2}
\end{table}

\section{Conclusions}
In this paper, we clarify the relation between the Noether charge associated to an arbitrary vector field and the equations of motions. This is unfortunately ignored in \cite{Liu:2017kml}, in which it was first proposed that in asymptotically AdS space-time, the Noether charge associated to a time-like Killing vector field is dual to the heat current in the boundary theory. The conjecture has passed several non-trivial tests in \cite{Liu:2017kml}. However, its validity for general holographic theories is relatively underexplored. Our new observation is the charge $\sqrt{-g}\mathcal{Q}^{rx_i}$ is nothing else but an integration constant of the linearized equations of motions. Furthermore, we find that the charge $\mathcal{Q}^{rt}$ may be interpreted as the thermal energy of the boundary theory so $\mathcal{Q}^{ra}$ is dual to the covariant thermal current in the boundary (see Eq.(\ref{covariantheat})). This is very similar to the case of charge density and electric current in the boundary. With these results, we argue that and the Noether charge may be viewed as an alternatively reasonable definition of the heat current for general holographic theories, in addition to the original one by using holographic stress tensor.

We also observe that the charge $\sqrt{-g}\mathcal{Q}^{rx_2}$ associated to a space-like Killing vector $\xi=\partial/\partial x_1$ is an integration constant of the Einstein equation $H^{x_1x_2}$ which determines the dynamics of the transverse perturbation $\delta g_{x_1x_2}(t\,,r)$. When the perturbation $\delta g_{x_1x_2}(t\,,r)$ depends linearly on time, the charge $\sqrt{-g}\mathcal{Q}^{rx_2}$ turns out to be radially conserved. We interpret the Noether charge $\mathcal{Q}^{rx_2}$ as shear stress of the dual fluid and propose a new method to compute the shear viscosity to entropy ratio by simply using the near horizon solutions. For Einstein gravity, our new method reproduces the celebrated result $1/4\pi$. For Gauss-Bonnet gravity, we find that the ratio $\eta/S$ receives corrections from the Gauss-Bonnet coupling as well as the argument of the scalar potential on the event horizon (see Eq.(\ref{shearGB})). For pure Gauss-Bonnet gravity, it reproduces the known result in \cite{shenker} but for general cases, it has never been found before. To give a cross check, we rederive the ratio $\eta/S$ in terms of asymptotic solutions by using the conventional approach. We numerically solve the linearized Einstein equation at small $\omega$ approximation and find that the numerical result is in good agreement with the analytical result Eq.(\ref{shearGB}) at sufficiently high precision.

\section*{Acknowledgments}
Z.Y. Fan thanks Wei-Jia Li, Jian-Pin Wu and Peng Liu for valuable discussions. This work is supported in part by the National Natural Science Foundations of China with Grant No. 11575270 and also supported by Guangdong Innovation Team for Astrophysics(2014KCXTD014).

\section{Appendix: Viscosity bound and causality violation}

To see whether and how the scalar potential affects viscosity bound and causality violation, we shall study the graviton wave equation in large momentum limit by following \cite{Brigante:2008gz}. To compare with the results there, we focus on $n=5$ dimension and will set $\alpha=\fft 12 \lambda_{GB}\,\ell^2$ when necessary. We take the transverse perturbation $\Psi=r^{-2}\delta g_{x_1x_2}$ to be independent of $x_{1\,,2}$ and write
\be \Psi(t\,,r\,,\vec{x})=e^{i k\cdot x}\phi_{en}(t\,,r\,,\vec x) \,,\ee
where $k=(-\omega\,,k_r\,,0\,,0\,,q)$ and $\phi_{en}$ denotes a slowly varying envelope function. At large momentum limit, the full graviton wave equation $H_{x_1x_2}$ gives at leading order
\be 0\simeq k^\mu k^\nu g_{\mu\nu}^{\mathrm{eff}} \,,\ee
where $k^\mu\equiv g^{\mu\nu}_{\mathrm{eff}} k_\nu$ and the effective geometry is
\be\label{graviton} g_{\mu\nu}^{\mathrm{eff}} dx^\mu dx^\nu=h(r)\Big(-dt^2+\fft{1}{c_g^2}\, dx_3^2\Big)+\fft{1}{\tilde{f}(r)}\,dr^2 \,,\quad\,\, \tilde{f}=\fft{r h-2\alpha f h'}{h\big(r-2\alpha f'\big)}\,f\,.\ee
In our numerics, we find that the function $\tilde{f}$ is a monotonically increasing function of $r$ and $\tilde{f}>0$ outside the black hole event horizon.
In the effective geometry, $c_g^2$ can be interpreted as local speed of graviton on a constant $r$-hypersurface, given by
\bea c_g^2&=&\fft{\ell^2 h}{r\big( r-2\alpha f' \big)}\Big(1-\alpha f\big( \ft{2h''}{h}-\ft{h'^2}{h^2}+\ft{h'f'}{h f} \big) \Big)  \nn\\
          &=&c_b^2 \big(1-2\alpha\ft{f'}{r} \big)^{-1}\Big(1-\alpha f\big( \ft{2h''}{h}-\ft{h'^2}{h^2}+\ft{h'f'}{h f} \big) \Big)\,,
\eea
where $c_b^2=\ell^2 h/r^2$ is the local speed of light defined by the background metric (\ref{static}). Thus the graviton cone in general does not coincide with the standard light cone defined by the background metric, as expected for a gravity theory with higher derivative terms. It was argued in \cite{shenker} that a graviton wave packet moving at speed $c_g$ in the bulk should be dual to the disturbances of the stress tensor propagating with the same velocity in the boundary. Hence, it is instructive to compare $c_g$ with the boundary speed of light. For later convenience, we properly scale the time coordinate such that the boundary speed of light is equal to unity. To see whether the local speed of graviton can be greater than unity for certain range of $r$, we first examine its behavior at infinity. We demand the metric is asymptotic to locally AdS space-time, in which the metric functions behave as
\be h=g^2r^2-\fft{2\mu_g}{r^2}+\cdots\,,\qquad f=\widetilde{g}^2 r^2+\cdots \,.\ee
We find
\be\label{asympcg} c_g^2=1+\fft{a_1}{r^4}+\cdots\,, \qquad a_1=-\fft{2\mu_g\ell^2(1-2\alpha \widetilde{g}^2)}{1-4\alpha\widetilde{g}^2} \,.\ee
It follows that when $\lambda_{GB}> 9/100$, $a_1$ will be positive and hence $c_g^2$ is greater than $1$. It should be emphasized that this has nothing to do with the presence of the scalar field.

The analysis in \cite{Brigante:2008gz} relates $c_g>1$ to microcausality violation in the boundary theory by studying a graviton null geodesic in the effective geometry (\ref{graviton})
\be 0\simeq \fft{dx^\mu}{ds}\fft{dx^\nu}{ds} g_{\mu\nu}^{\mathrm{eff}} \,,\ee
with the identification $dx^\mu/ds\equiv k^\mu$. This is valid at large momentum limit. Although we have slightly different formulas, we find that the analysis which followed in \cite{Brigante:2008gz} still holds for our cases. So we will drop the details in the following. The readers who have interests should refer to \cite{Brigante:2008gz}.

\begin{figure}[ht]
\includegraphics[width=210pt]{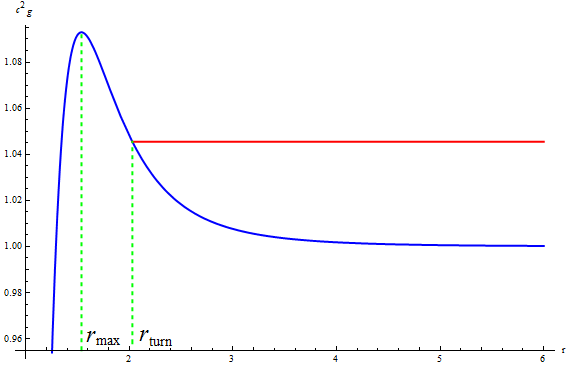}
\includegraphics[width=210pt]{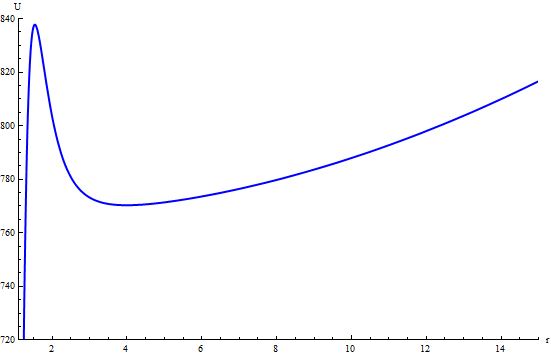}
\caption{{\it The left panel is $c_g^2(r)$ as a function of $r$ for $\lambda_{GB}=0.1>9/100$. $c_g^2$ has a local maximum at $r_{\mathrm{max}}$. The horizontal line denotes the trajectory of a bouncing graviton null geodesic with the turning point $r_{\mathrm{turn}}$. The right panel is the Schr\"{o}dinger potential $U(r)$ as a function of $r$ for the same $\lambda_{GB}$ and $q=100$. For large enough $q$, $U(r)$ develops a well and admits metastable states which correspond to quasiparticles in the boundary theory. }}
\label{cas1}\end{figure}
To illustrate the major results in \cite{Brigante:2008gz}, we consider the case in which $c_g(r)$ has a local maximum greater than $1$, for example when $\lambda_{GB}>9/100$. One finds that there exists a bouncing null geodesic which starts and ends at the boundary (see Fig.\ref{cas1}).
The turning point $r_{\mathrm{turn}}(\varepsilon)$ is fixed by the condition $\varepsilon^2=c_g^2(r_{\mathrm{turn}})$, where $\varepsilon\equiv \omega/q$. We have
\bea
&&\Delta t(\varepsilon)=2\int_{r_{\mathrm{turn}}(\varepsilon)}^\infty \fft{\varepsilon}{p(r)\sqrt{\varepsilon^2-c_g^2}}\mathrm{d}r \,,\nn\\
&&\Delta x_3(\varepsilon)=2\int_{r_{\mathrm{turn}}(\varepsilon)}^\infty \fft{c_g^2}{p(r)\sqrt{\varepsilon^2-c_g^2}}\mathrm{d}r\,,
\eea
where $p(r)=\sqrt{h\tilde{f}}$. As $r_{\mathrm{turn}}\rightarrow r_{\mathrm{max}}$, $\varepsilon\rightarrow c_{g\,,\mathrm{max}}$, the geodesic hovers near $r_{\mathrm{max}}$ for a long time with a propagating velocity $c_{g\,,\mathrm{max}}$ along the $x_3$ direction. This is easily seen since the above integrals are dominated by contributions near $r_{\mathrm{max}}$. Thus, one finds $\fft{\Delta x_3(\varepsilon)}{\Delta t(\varepsilon)}\rightarrow c_{g\,,\mathrm{max}}>1$. By rewriting the full graviton wave equation in a Schr\"{o}dinger form,\footnote{
In the general static background (\ref{static}), we have
\bea
&&-\partial_y^2 \phi+U(y)\phi=\omega^2 \phi\,,\quad \phi=B \Psi e^{-i k\cdot x}\,,\quad \fft{dy}{dr}\equiv p(r)  \,,\nn\\
&& U(y)=q^2 c_g^2+U_1\,,\quad  U_1=\fft{\partial^2_y B}{B}\,,\quad \fft{\partial_y B}{B}\equiv W\,,
\eea
where the function $W$ is given by
\be W=\fft{2q(r)-p'(r)}{4\sqrt{p(r)}} \,,\quad q(r)=(r-2\alpha f')^{-1}\Big(3hf+\ft 12 r \big( hf\big)'-\alpha f^2h'\big(\ft{2h''}{h}-\ft{h'}{h}+\ft{3f'}{f}+\ft{4}{r} \big) \Big)\,.\ee
Here the prime denotes derivative with respect to $r$.
}
it was further shown in \cite{Brigante:2008gz} that the superluminal propagation of graviton in the bulk is dual to the superluminal propagation of metastable quasiparticles in the boundary with $\fft{\Delta x_3}{\Delta t}$ identified as group velocity of the quasiparticles. Therefore, the fact $c_{g\,,\mathrm{max}}>1$ signals microcausality violation in the boundary theory, rendering the theory inconsistent. The crucial result of \cite{Brigante:2008gz} is for any $\lambda_{GB}\leq 9/100$, $c_g^2$ cannot be greater than $1$ any longer. Hence, it leads to a modified lower bound on viscosity $\eta/S\geq \big(1/4\pi)\times \big(16/25\big)$ for Gauss-Bonnet black holes (we call it ``B-bound" in the following). This is a concrete example which shows how the viscosity bound is correlated to the inconsistency of a theory.
\begin{figure}[ht]
\centering
\includegraphics[width=140pt]{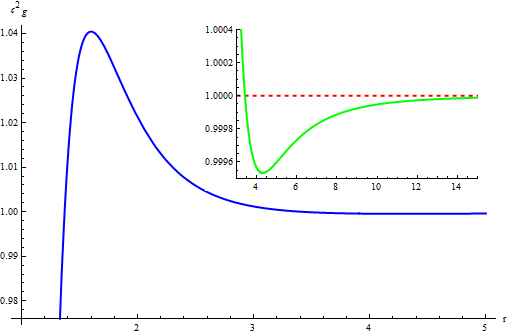}
\includegraphics[width=140pt]{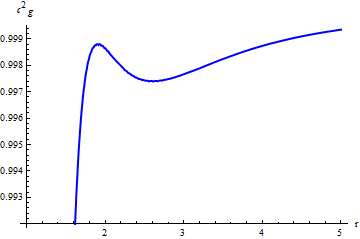}
\includegraphics[width=140pt]{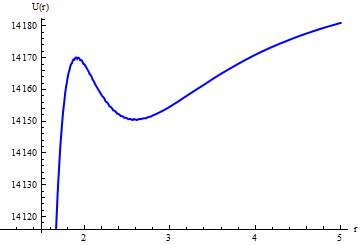}
\caption{{\it The left panel is $c_g^2$ as a function of $r$ for $\lambda_{GB}=6/100$ and $\phi_0=3>\phi_c\simeq 2.62$. In this case, the local maximum $c_{g\,,\mathrm{max}}$ is greater than $1$, implying microcausality violation in the boundary theory, in spite of that $c_g^2$ approaches to unity from below at infinity. The middle and right panels are $c_g^2$ and the Schr\"{o}dinger potential $U(r)$ for the same $\lambda_{GB}$ and $\phi_0=2.6<\phi_c$. In this case, $c_g$ still has a local maximum but $c_{g\,,\mathrm{max}}<1$, implying the existence of new quasiparticles in the boundary theory, as easily seen from the Schr\"{o}dinger potential. We have set $m^2=-4\ell^2\,,\ell=1$.}}
\label{cas2}\end{figure}

We do not find any indication that disagrees with the results in \cite{Brigante:2008gz}. However, our new observation is in the presence of a scalar field, the asymptotic analysis in (\ref{asympcg}) is not sufficient to constrain the Gauss-Bonnet coupling as well as the lower bound on viscosity. To be concrete, we focus on a free massive scalar field and we are interested in the parameters space $\alpha>0\,,m_{BF}^2\leq m^2<0$, in which the KSS bound is violated. We find that for each given $\lambda_{GB}<9/100$, there exists a critical value $\phi_c$ for $\phi_0$ such that when $\phi_0>\phi_c$, $c_g^2$ still has a local maximum that is greater than $1$ (see Fig.\ref{cas2}), although it approaches to unity from below at infinity. Thus, causality requires $\phi_0\leq \phi_c$ for any given $\lambda_{GB}<9/100$.

Interestingly, we find that when $\phi_0$ close to $\phi_c$ from below, there exists a tiny interval in which $c_g$ still has a local maximum which is however smaller than $1$. For example, in Fig.\ref{cas2}, we also plot $c_g^2$ and the Schr\"{o}dinger potential for $\lambda_{GB}=6/100\,,m^2=-4\ell^2$ with $\phi_0=2.6$ slightly smaller than the critical value $\phi_c\simeq 2.62$. The behavior of both implies the existence of new quasiparticles which might be stable in the boundary theory. This is a new result that was not obtained in \cite{Brigante:2008gz} for pure Gauss-Bonnet gravity.
\begin{figure}[ht]
\includegraphics[width=210pt]{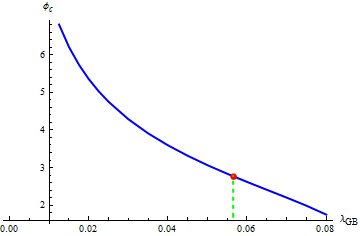}
\includegraphics[width=210pt]{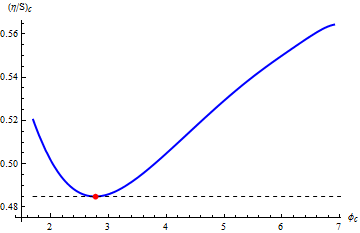}
\caption{{\it The left panel is the critical value $\phi_c$ as a function of $\lambda_{GB}$ for $m^2=-4\ell^2$. The right panel is the critical ratio $\eta/S$ (measured in units of $1/4\pi$) as a function of $\phi_c$. The local minimum of the viscosity is $\eta/S\simeq 1/4\pi\times 0.4848$ at $\phi_c\simeq 2.7714$, corresponding to $\lambda_{GB}\simeq 0.0565$.  }}
\label{cas3}\end{figure}

Here comes an intriguing question: with a scalar field, does there exist a new bound on viscosity that is similar to the B-bound proposed in \cite{Brigante:2008gz}?. We scan the parameters space and find the answer is yes. To explain this, we first rewrite our result (\ref{shearGB}) as follows
\bea
\fft{\eta}{S}&=&\fft{1}{4\pi}\Big(1-4\lambda_{GB}-\ft 16 |m^2|\ell^2 \,\lambda_{GB}\,\phi_0^2 \Big) \nn\\
&\geq&\fft{1}{4\pi}\Big(1-4\lambda_{GB}-\ft 16 |m^2|\ell^2\,\lambda_{GB}\, \phi_c^2(\lambda_{GB}) \Big)\nn\\
&\equiv& \Big( \fft{\eta}{S}\Big)_c
\eea
It follows that the answer of above question highly depends on the functional relation $\phi_c=\phi_c(\lambda_{GB})$. Numerically, we find that $\phi_c$ is a monotonically decreasing function of $\lambda_{GB}$. However, it turns out that the critical ratio $\big(\eta/S\big)_c$ as a function of $\phi_c$ indeed has a local minimum, as shown in Fig.\ref{cas3}. For example, for $m^2=-4\ell^2$, we find $\big(\eta/S\big)_{c\,,min}\simeq 1/4\pi\times 0.4848 $ at $\phi_c\simeq 2.7714\,,\lambda_{GB}\simeq 0.0565$ and for $m^2=-2\ell^2$, $\big(\eta/S\big)_{c\,,min}\simeq 1/4\pi\times 0.5532 $ at $\phi_c\simeq 3.7457\,,\lambda_{GB}\simeq 0.0513$. These lead to generalized B-bound for Gauss-Bonnet black holes with scalar hair. Note that the new bound is smaller than the B-bound $1/4\pi \times 0.64$ and $\big(\eta/S\big)_{c\,,min}$ increases as $|m^2|$ decreases. Hence, we naturally expect that our new bound smoothly approaches to the B-bound in the limit $m^2\rightarrow 0$.

To end this section, we argue that the presence of a scalar field tends to lower the bound further on viscosity. It is also interesting to study whether this is the case for a scalar field with self-interactions. Finally, it is worth emphasizing that our results also support the idea that the consistency of a theory is correlated to the viscosity bound.


\begin{thebibliography}{100}

\bibitem{Buchel:2003tz}
  A.~Buchel and J.~T.~Liu,
{\it Universality of the shear viscosity in supergravity,}
  Phys.\ Rev.\ Lett.\  {\bf 93}, 090602 (2004),
  hep-th/0311175.

\bibitem{Buchel:2004qq}
  A.~Buchel,
{\it On universality of stress-energy tensor correlation functions in
supergravity,}
  Phys.\ Lett.\ B {\bf 609}, 392 (2005),
  hep-th/0408095.

\bibitem{Benincasa:2006fu}
  P.~Benincasa, A.~Buchel and R.~Naryshkin,
{\it The shear viscosity of gauge theory plasma with chemical potentials,}
  Phys.\ Lett.\ B {\bf 645}, 309 (2007),
  hep-th/0610145.

\bibitem{Landsteiner:2007bd}
  K.~Landsteiner and J.~Mas,
{\it The shear viscosity of the non-commutative plasma,}
  JHEP {\bf 0707}, 088 (2007)
  arXiv:0706.0411 [hep-th].

\bibitem{Cremonini:2011iq}
  S.~Cremonini,
{\it The shear viscosity to entropy ratio: A status report,}
  Mod.\ Phys.\ Lett.\ B {\bf 25}, 1867 (2011)
  arXiv:1108.0677 [hep-th].

\bibitem{Iqbal:2008by}
  N.~Iqbal and H.~Liu,
{\it Universality of the hydrodynamic limit in AdS/CFT and
the membrane paradigm,}
  Phys.\ Rev.\ D {\bf 79}, 025023 (2009)
  arXiv:0809.3808 [hep-th].

\bibitem{Cai:2008ph}
  R.G.~Cai, Z.Y.~Nie and Y.W.~Sun,
{\it Shear viscosity from effective couplings of gravitons,}
  Phys.\ Rev.\ D {\bf 78}, 126007 (2008)
  [arXiv:0811.1665 [hep-th]].

\bibitem{Cai:2009zv}
  R.G.~Cai, Z.Y.~Nie, N.~Ohta and Y.W.~Sun,
{\it Shear viscosity from Gauss-Bonnet gravity with a dilaton coupling,}
  Phys.\ Rev.\ D {\bf 79}, 066004 (2009)
  [arXiv:0901.1421 [hep-th]].

\bibitem{Brustein:2007jj}
  R.~Brustein, D.~Gorbonos and M.~Hadad,
{\it Wald's entropy is equal to a quarter of the horizon area in units of the effective gravitational coupling,}
  Phys.\ Rev.\ D {\bf 79}, 044025 (2009)
  [arXiv:0712.3206 [hep-th]].

\bibitem{Kats:2007mq}
  Y.~Kats and P.~Petrov,
{\it Effect of curvature squared corrections in AdS on the viscosity of the dual gauge theory,}
  JHEP {\bf 0901}, 044 (2009)
  [arXiv:0712.0743 [hep-th]].

\bibitem{shenker} M. Brigante, H. Liu, R.C. Myers, S. Shenker and S. Yaida,
{\it Viscosity bound violation in higher derivative gravity},
Phys.\ Rev.\ D {\bf 77}, 126006 (2008) [arXiv:0712.0805 [hep-th]].

\bibitem{Natsuume:2010ky}
  M.~Natsuume and M.~Ohta,
{\it The shear viscosity of holographic superfluids,}
  Prog.\ Theor.\ Phys.\  {\bf 124}, 931 (2010)
  [arXiv:1008.4142 [hep-th]].

\bibitem{Erdmenger:2010xm}
  J.~Erdmenger, P.~Kerner and H.~Zeller,
{\it Non-universal shear viscosity from Einstein gravity,}
  Phys.\ Lett.\ B {\bf 699}, 301 (2011)
  [arXiv:1011.5912 [hep-th]].

\bibitem{Ovdat:2014ipa}
  O.~Ovdat and A.~Yarom,
{\it A modulated shear to entropy ratio,}
  JHEP {\bf 1411}, 019 (2014)
  [arXiv:1407.6372 [hep-th]].

\bibitem{Ge:2014aza}
  X.~H.~Ge, Y.~Ling, C.~Niu and S.~J.~Sin,
  {\it Thermoelectric conductivities, shear viscosity, and stability in an anisotropic linear axion model,}
  Phys.\ Rev.\ D {\bf 92}, no. 10, 106005 (2015)
  [arXiv:1412.8346 [hep-th]].

\bibitem{Liu:2015tqa}
  H.~S.~Liu, H.~Lu and C.~N.~Pope,
  {\it Generalized Smarr formula and the viscosity bound for Einstein-Maxwell-dilaton black holes,}
  Phys.\ Rev.\ D {\bf 92}, 064014 (2015) [arXiv:1507.02294 [hep-th]].


\bibitem{Sadeghi:2015vaa}
  M.~Sadeghi and S.~Parvizi,
  {\it Hydrodynamics of a black brane in Gauss¨CBonnet massive gravity,}
  Class.\ Quant.\ Grav.\  {\bf 33}, no. 3, 035005 (2016)
  [arXiv:1507.07183 [hep-th]].

\bibitem{Parvizi:2017boc}
  S.~Parvizi and M.~Sadeghi,
  {\it Holographic Aspects of a Higher Curvature Massive Gravity,}
  arXiv:1704.00441 [hep-th].


\bibitem{Hartnoll:2009sz}
  S.~A.~Hartnoll,
  {\it Lectures on holographic methods for condensed matter physics,}
  Class.\ Quant.\ Grav.\  {\bf 26}, 224002 (2009)
  [arXiv:0903.3246 [hep-th]].


\bibitem{Herzog:2009xv}
  C.~P.~Herzog,
  {\it Lectures on Holographic Superfluidity and Superconductivity,}
  J.\ Phys.\ A {\bf 42}, 343001 (2009)
  [arXiv:0904.1975 [hep-th]].


\bibitem{Hartnoll:2016apf}
  S.~A.~Hartnoll, A.~Lucas and S.~Sachdev,
  {\it Holographic quantum matter,}
  arXiv:1612.07324 [hep-th].


\bibitem{sonsta} D.T. Son and A.O. Starinets,
{\it Minkowski space correlators in AdS/CFT correspondence:
Recipe and applications},
JHEP {\bf 0209}, 042 (2002), hep-th/0205051.



\bibitem{Policastro:2001yc}
  G.~Policastro, D.T.~Son and A.~O.~Starinets,
{\it The shear viscosity of strongly coupled ${\cal N}=4$
supersymmetric Yang-Mills plasma,}
  Phys.\ Rev.\ Lett.\  {\bf 87}, 081601 (2001),
 hep-th/0104066.

\bibitem{KSS} P. Kovtun, D.T. Son and A.O. Starinets,
{\it Holography and hydrodynamics: Diffusion on stretched horizons},
JHEP {\bf 0310}, 064 (2003), hep-th/0309213.

\bibitem{KSS0} P.~Kovtun, D.T.~Son and A.O.~Starinets,
{\it Viscosity in strongly interacting quantum field theories from black
hole physics},  Phys.\ Rev.\ Lett.\  {\bf 94}, 111601 (2005),
hep-th/0405231.


\bibitem{Lucas:2015vna}
  A.~Lucas,
  {\it Conductivity of a strange metal: from holography to memory functions,}
  JHEP {\bf 1503}, 071 (2015)
  [arXiv:1501.05656 [hep-th]].

\bibitem{Donos:2014uba}
  A.~Donos and J.~P.~Gauntlett,
  { \it Novel metals and insulators from holography,}
  JHEP {\bf 1406}, 007 (2014)
  [arXiv:1401.5077 [hep-th]].

\bibitem{Donos:2014cya}
  A.~Donos and J.~P.~Gauntlett,
  {\it Thermoelectric DC conductivities from black hole horizons,}
  JHEP {\bf 1411}, 081 (2014)
  [arXiv:1406.4742 [hep-th]].

\bibitem{Liu:2017kml}
  H.~S.~Liu, H.~Lu and C.~N.~Pope,
  {\it Holographic Heat Current as Noether Current,}
  JHEP {\bf 1709}, 146 (2017)
  [arXiv:1708.02329 [hep-th]].



\bibitem{wald1}
  R.M.~Wald,
{\it Black hole entropy is the Noether charge},
Phys.\ Rev.\ D {\bf 48}, 3427 (1993), gr-qc/9307038.

\bibitem{wald2}  V.~Iyer and R.M.~Wald,
{\it Some properties of Noether charge and a proposal for
dynamical black hole entropy,}
Phys.\ Rev.\ D {\bf 50}, 846 (1994),  gr-qc/9403028.


\bibitem{Papadimitriou:2005ii}
  I.~Papadimitriou and K.~Skenderis,
  {\it Thermodynamics of asymptotically locally AdS spacetimes,}
  JHEP {\bf 0508}, 004 (2005)
  [hep-th/0505190].



\bibitem{Lu:2013ura}
  H.~L¨¹, Y.~Pang and C.~N.~Pope,
  {\it AdS Dyonic Black Hole and its Thermodynamics,}
  JHEP {\bf 1311}, 033 (2013)
  [arXiv:1307.6243 [hep-th]].

\bibitem{Liu:2013gja}
  H.~S.~Liu and H.~L¨¹,
  {\it Scalar Charges in Asymptotic AdS Geometries,}
  Phys.\ Lett.\ B {\bf 730}, 267 (2014),
  [arXiv:1401.0010 [hep-th]].

\bibitem{Liu:2014tra}
  H.~S.~Liu, H.~L¨¹ and C.~N.~Pope,
  {\it Thermodynamics of Einstein-Proca AdS Black Holes,}
  JHEP {\bf 1406}, 109 (2014)
  [arXiv:1402.5153 [hep-th]].

\bibitem{Liu:2014dva}
  H.~S.~Liu and H.~L¨¹,
  {\it Thermodynamics of Lifshitz Black Holes,}
  JHEP {\bf 1412}, 071 (2014)


\bibitem{Fan:2014ixa}
  Z.~Y.~Fan and H.~L¨¹,
  {\it SU(2)-Colored (A)dS Black Holes in Conformal Gravity,}
  JHEP {\bf 1502}, 013 (2015)
  [arXiv:1411.5372 [hep-th]].

\bibitem{Lu:2014maa}
  H.~Lu, C.~N.~Pope and Q.~Wen,
  {\it Thermodynamics of AdS Black Holes in Einstein-Scalar Gravity,}
  JHEP {\bf 1503}, 165 (2015)
  [arXiv:1408.1514 [hep-th]].

\bibitem{Fan:2014ala}
  Z.~Y.~Fan and H.~Lu,
  {\it Thermodynamical First Laws of Black Holes in Quadratically-Extended Gravities,}
  Phys.\ Rev.\ D {\bf 91}, no. 6, 064009 (2015)
  [arXiv:1501.00006 [hep-th]].


\bibitem{Fan:2015yza}
  Z.~Y.~Fan and H.~Lu,
  {\it Charged Black Holes in Colored Lifshitz Spacetimes,}
  Phys.\ Lett.\ B {\bf 743}, 290 (2015)
  [arXiv:1501.01727 [hep-th]].


\bibitem{Chen:2016qks}
  B.~Chen, Z.~Y.~Fan and L.~Y.~Zhu,
  {\it AdS and Lifshitz Scalar Hairy Black Holes in Gauss-Bonnet Gravity,}
  Phys.\ Rev.\ D {\bf 94}, no. 6, 064005 (2016)
  [arXiv:1604.08282 [hep-th]].



\bibitem{Feng:2015oea}
  X.~H.~Feng, H.~S.~Liu, H.~L¨¹ and C.~N.~Pope,
  {\it Black Hole Entropy and Viscosity Bound in Horndeski Gravity,}
  JHEP {\bf 1511}, 176 (2015)
  [arXiv:1509.07142 [hep-th]].


\bibitem{Feng:2015wvb}
  X.~H.~Feng, H.~S.~Liu, H.~L¨¹ and C.~N.~Pope,
  {\it Thermodynamics of Charged Black Holes in Einstein-Horndeski-Maxwell Theory,}
  Phys.\ Rev.\ D {\bf 93}, no. 4, 044030 (2016),
  [arXiv:1512.02659 [hep-th]].


\bibitem{Fan:2016jnz}
  Z.~Y.~Fan,
  {\it Black holes with vector hair,}
  JHEP {\bf 1609}, 039 (2016)
  [arXiv:1606.00684 [hep-th]].


\bibitem{Fan:2017bka}
  Z.~Y.~Fan,
  {\it Black holes in vector-tensor theories and their thermodynamics,}
  arXiv:1709.04392 [hep-th].

\bibitem{Brigante:2008gz}
  M.~Brigante, H.~Liu, R.~C.~Myers, S.~Shenker and S.~Yaida,
  {\it The Viscosity Bound and Causality Violation,}
  Phys.\ Rev.\ Lett.\  {\bf 100}, 191601 (2008)
  [arXiv:0802.3318 [hep-th]].







\end{thebibliography}
\end{document}